\documentclass[aps, prb, amssymb, amsmath, superscriptaddress, 
twocolumn] {revtex4-2}

\usepackage{chngcntr}

\usepackage{graphicx}
\usepackage{color}
\usepackage{amsmath}
\usepackage{enumitem}
\usepackage{amssymb}
\usepackage{hyperref}
\usepackage{cancel}
\usepackage{ulem}
\usepackage{multirow}

\newcommand{\be}{\begin{equation}}
	\newcommand{\ee}{\end{equation}}

\newcommand{\bea}{\begin{eqnarray}}
	\newcommand{\eea}{\end{eqnarray}}

\newcommand{\lb}{\left[}
\newcommand{\rb}{\right]}
\newcommand{\lp}{\left(}
\newcommand{\rp}{\right)}
\newcommand{\sgn}{{\rm sgn}}

\renewcommand{\vec}[1]{{\boldsymbol #1}}

\newcommand{\addLL}[1]{\textcolor{black}{#1}}
\newcommand{\addQ}[1]{\textcolor{black}{#1}}

\makeatother

\begin{document}
\title{Charge and spin density wave orders in field-biased Bernal bilayer graphene}
\author{Zhiyu Dong}
\affiliation{Department of Physics, Massachusetts Institute of Technology, Cambridge, MA 02139}
\affiliation{Department of Physics and Institute for Quantum Information and Matter, California Institute of Technology, Pasadena, California 91125}
\author{Patrick A. Lee}
\affiliation{Department of Physics, Massachusetts Institute of Technology, Cambridge, MA 02139}
\author{Leonid Levitov}
\affiliation{Department of Physics, Massachusetts Institute of Technology, Cambridge, MA 02139}

\begin{abstract}
 This paper aims to clarify the nature of a surprising ordered phase recently reported in biased Bernal bilayer graphene that occurs at the phase boundary between the isospin-polarized and unpolarized phases. Strong nonlinearity of transport at abnormally small currents, with $dI/dV$ vs. $I$ sharply rising and then falling back, is typical for a charge/spin-density-wave state (CDW or SDW) sliding transport. Here, however, it is observed at an isospin-order phase boundary, prompting a question about the CDW/SDW mechanism and its relation to the quantum critical point. We argue that the observed phase diagram cannot be understood within a standard weak-coupling picture. Rather, it points to a mechanism that relies on an effective interaction enhancement at a quantum critical point. We develop a detailed strong-coupling framework accounting for the soft collective modes that explain these observations.
\end{abstract}

\maketitle

Graphene-based materials provide new opportunities for studying exotic orders driven by electron-electron interaction. One system that received much attention is the twisted bilayer graphene\cite{andrei2020graphene}, where the flattened electron bands lead to the dominant role of interaction\cite{Bistritzer12233}, giving rise to a variety of correlated insulating and superconducting phases\cite{cao2018insulator, cao2021nematicity,jaoui2022quantum,cao2018SC,lu2019superconductors,stepanov2020untying,saito2020independent,oh2021evidence,liu2021tuning}.
Another system that came to light recently is the nontwisted multilayer graphene\cite{zhou2021half, zhou2021superconductivity,zhou2022isospin,de2021cascade,seiler2021quantum}. When biased by a transverse electric field, a gap at charge neutrality opens up, a transformation that flattens the electron dispersion\cite{mccann2013electronic}. The flattened bands reduce the kinetic energy and boost the effects of interactions, leading to interesting strongly correlated phases\cite{zhou2021half, zhou2021superconductivity,zhou2022isospin,de2021cascade,seiler2021quantum}. Among the phases where the order type has been identified is a cascade of isospin-polarized phases and adjacent superconducting phases 
 \cite{zhou2022isospin,de2021cascade,seiler2021quantum}.


In addition, in some of the observed phases, the order type has not yet been identified. 
Interestingly, near one of the isospin phase transitions seen in Bernal bilayer graphene (BBG), the system exhibits a rise in resistivity below about 50 mK, together with nonlinear resistivity\cite{zhou2022isospin}.  The differential resistivity $dV/dI$ starts at a constant value at a small current, and abruptly drops around a threshold current of $\sim 5\rm{nA}$, and recovers the high-temperature value at a 
current of $\sim 10\rm{nA}$ (see Fig 5a). This has led the authors to propose an interpretation in terms of the charge or spin density wave (CDW/SDW) and its sliding motion due to depinning \cite{fleming1979sliding}. These data pose the question of the mechanism of this conjectured density-wave phase and why it is associated with the phase boundary of isospin (valley) polarization.

The answer to this question is nontrivial because the standard weak-coupling scenario for CDW/SDW cannot explain the measured phase diagram. Indeed, the CDW/SDW order predicted in this way would be insensitive to the proximity to the onset of isospin polarization: the $T_c$ in a  weak-coupling theory should not exhibit an abrupt enhancement at the isospin phase transition. This is discussed in detail in Sec.\ref{sec:weak-coupling}.

\begin{figure}
	\includegraphics[width=1.0\columnwidth]{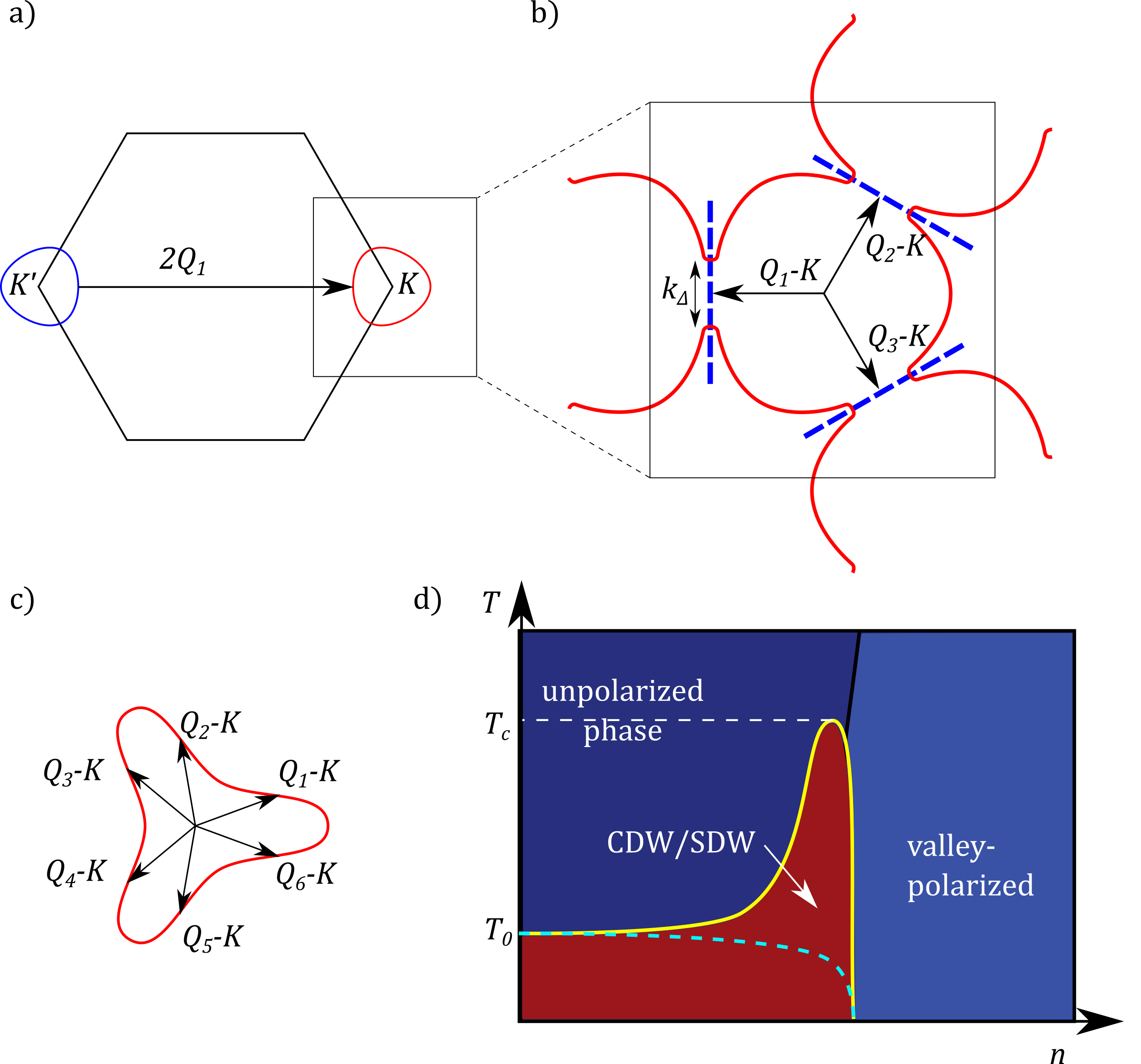}
	\centering
	\caption{ 
		a) The Fermi surface in the absence of interaction. The red and blue contour corresponds to Fermi surfaces in valley $K$ and $K'$ respectively. $Q_{1}$, $Q_{2}$, and $Q_{3}$ are the maxima of the radius of curvature on the Fermi surface where CDW/SDW gap tends to open. In this case, we obtain a triple-Q CDW/SDW instability.
		b) The case of concave Fermi surface, where the curvature has six minima. In general, this situation might occur but, for simplicity, we only focus on the triple-Q CDW/SDW in this paper. 	Dashed blue lines are the edges of the mini Brillouin zone where the gap opening distorts the Fermi surface. 
c) The Fermi surface in the presence of a triple-Q CDW/SDW. Gaps open at three points in each valley. 
d) The schematic finite-temperature phase diagram. The yellow line and the cyan dashed line are respectively the boundaries of the CDW phase predicted by strong-coupling theory and weak-coupling theory. 
		The strong coupling theory predicts a significant enhancement of CDW critical temperature at the valley-polarizing criticality $T_c\gg T_0$. When temperature lies between $T_c$ and $T_0$, we obtain a phase diagram resembling measured one \cite{zhou2022isospin}.   }\label{fig:FS}
\end{figure}

\begin{figure}
	\includegraphics[width=1.0\columnwidth]{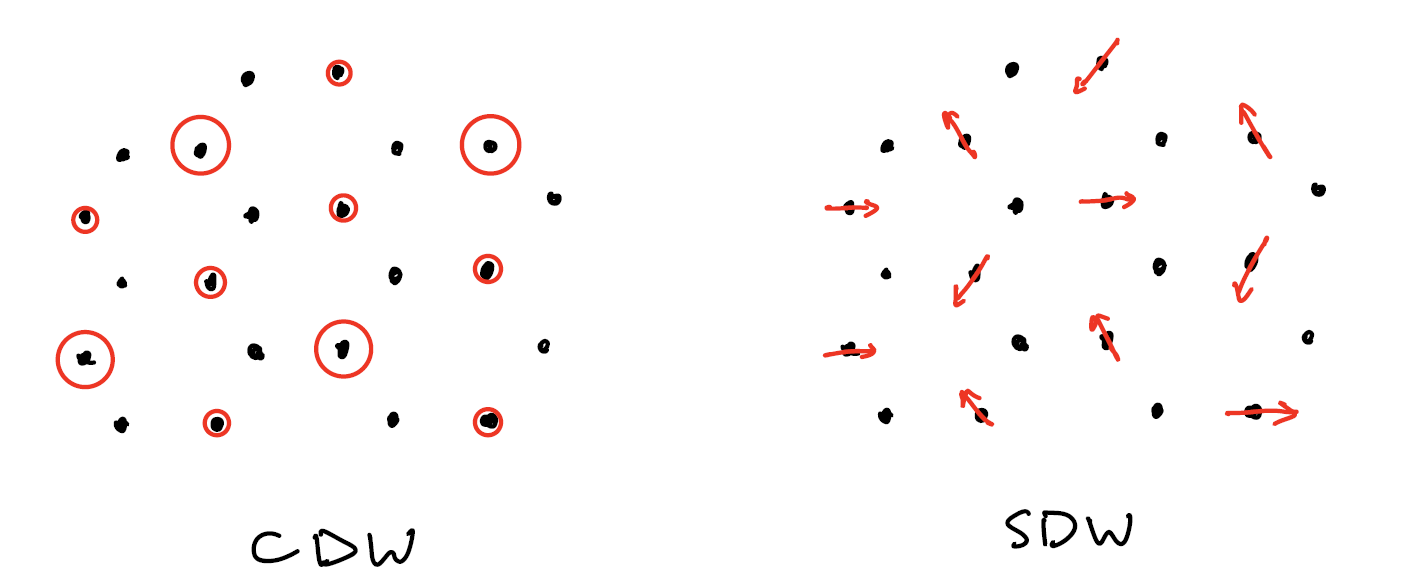}
	\centering
	\caption{
The predicted CDW and SDW patterns in real space. Black dots represent carbon atoms 
of the top layer of Bernal bilayer graphene to which carriers are shifted upon application of a transverse field. Carriers predominantly populate one sublattice of the top layer, referred to as A, in which sites are not aligned with the sites at the bottom layer. Circles in a) and arrows in b) illustrate CDW and SDW order, respectively. Circles of different radii denote different local charge density. Different arrow orientations denote different local spin polarization. Sites of the other sublattice (B) support states with strong interlayer hybridization, which form a higher-energy band that does not play a role in electronic ordering    
Both CDW and SDW orders feature spatial modulation with periodicity set by the Fermi wavelength. 
Notably, they exhibit Kekul\'e-like atomic-scale patterns that triple the unit cell. Accordingly, in the SDW order, the spin density wave takes the form of a $120^{\circ}$ order in which spins on three different sublattices are oriented at $120^{\circ}$ with respect to each other.  }\label{fig:pattern}
\end{figure}

To the contrary, a strong-coupling approach 
accounting for the e-e interaction enhanced at the quantum-critical point offers a natural framework to understand the experimental findings. 
The key assumption is that the observed isospin order phase transition is either continuous or weakly first-order. 
Starting with this assumption, we predict a strong instability in the CDW and SDW channels. 
Specifically, we demonstrate that the interaction enhanced by the soft critical modes results in an abrupt rise of $T_c$ for the CDW/SDW order at the onset of isospin polarization. This prediction is in agreement with the observations (see Fig.\eqref{fig:FS} d)). 
The predicted CDW/SDW phase has several unique features.  First, it is a metallic state since the gaps only open in a segment of the Fermi surface. 
This is distinct from the CDW in 1D which gaps out the entire Fermi surface\cite{peierls1955quantum}. As a result, the Drude weight of the sliding mode in our system, which is proportional to the length of the gapped segment, is much smaller as compared to the 1D case \cite{lee1974conductivity}.  Moreover, here CDW/SDW takes a triple-$Q$ form, with the density-wave order parameter 
\bea
\text{for CDW:}\quad \rho(r) \sim \sum_{j=1,2,3} \rho_j \cos(2Q_jr+\phi_j). \\
\text{for SDW:}\quad \vec S(r) \sim \sum_{j=1,2,3} \vec S_j \cos(2Q_jr+\phi_j). 
\eea
The three $Q_j$ vectors are given by three minima of curvature on the Fermi surface [see Fig.\ref{fig:FS} a)], which are related by the $C_3$ rotation symmetry. In our theory, the three CDW amplitudes $\rho_{1}$, $ \rho_{2}$, and $\rho_{3}$ are equal, whereas the phases are decoupled at leading order in interaction. The same applies to SDW amplitudes and phases. 
The relation between phases $\phi_j$ 
in the ground state depends on subleading interactions. The most likely ground-state configuration is when the three components respect the $C_3$ rotation symmetry,  i.e. $\rho_1=\rho_2=\rho_3$ or $\vec S_1 = C_3 \vec S_{2} = C_3^{-1} \vec S_3$. The real-space patterns for these possible ground states are illustrated in Fig.\ref{fig:pattern}. In these states, the wavevectors $Q_j$ form a triade around $2K$, each $Q_j$ deviating from $2K$ by 
a small correction of order $k_F$. 
The CDW and SDW order parameters correspond to atomic-scale patterns that triple the unit cell, forming a $\sqrt{3}\times \sqrt{3}$ Kekul\'e-like charge-density-wave pattern and a similar $120^{\circ}$ spin-density-wave pattern as shown in Fig.\ref{fig:pattern}. These orders are similar to the orders recently seen in twisted bilayer graphene\cite{PhysRevLett.129.117602}. Because of the $\sim k_F$ corrections to the CDW/SDW wavevector lengths, these patterns are 
modulated with a much longer spatial periodicity 
set by the Fermi wavelength.


Our analysis will proceed as follows. First, after introducing a model of Bernal bilayer graphene with Coulomb interactions, we consider the weak-coupling 
picture of the CDW/SDW order 
and discuss its limitations. 
Next, we develop a strong-coupling framework and show that the CDW/SDW susceptibility is significantly enhanced at the isospin phase transition. Finally, we argue that the CDW/SDW predicted in this framework can explain the dependence of differential resistivity $dV/dI$ vs. $I$ reported in Ref.\cite{zhou2021superconductivity}. 
We discuss two possible explanations. One is the conventional scenario of  CDW/SDW sliding. 
Another explanation relies on Landau-Zener tunneling processes 
occurring in the CDW/SDW phase. Such processes may significantly impact the $V$-$I$ dependence 
provided the CDW/SDW gap is small. 
We find that both scenario 
can reproduce the correct orders of magnitude of the measured quantities.

\section{model}\label{sec:model}


Bernal bilayer graphene bands are conventionally described by quadratic Dirac bands with trigonal warping. However,
at low carrier densities typical for isospin and CDW orders 
the system can be described by a
one-band Hamiltonian with a short-range interaction: $H = H_0+H_{\rm int}$, 
\begin{align}
&H_0 = \sum_{\xi,s} \epsilon_{\xi }(\vec p) \psi_{\xi,s,\vec p}^{\dagger} \psi_{\xi,s,\vec p}\label{eq:H0}\\
\label{eq:H_int}
&H_{\rm int} = \frac{U}{2}\sum_{\xi\xi'ss'\vec p\vec p' \vec q}  \psi_{\xi,s,\vec p+\vec q}^\dagger \psi_{\xi',s',\vec p'- \vec q}^\dagger \psi_{\xi',s',\vec p'}\psi_{\xi,s,\vec p}
\end{align}
where $\xi,\xi'$ label valleys $K$ and $K'$,
$s,s'=\uparrow,\downarrow$ label spin. Fermionic variables $\psi_{\xi,s}$ represent the spin-$s$ electrons in the  conduction band in valleys $\xi=K,\,K'$. 
In what follows we will refer to the indices $\xi$ and $s$ as isospin indices.
The quantity $\epsilon_\xi(\vec p)$ describes the dispersion in valleys $\xi=K,\,K'$, where momentum $p$ is measured from $K$ and $K'$ points, respectively. Since here we are concerned with CDW order, we 
focus on the specific points on the Fermi surface where CDW nesting occurs. The dispersion at these points will be discussed shortly 
[see Eq.\eqref{eq:hotspot_energy}].

The interaction term, Eq.\eqref{eq:H_int}, describes electron-electron 
repulsion potential projected onto the BBG conduction bands. Here, we ignore the intervalley scattering ($K\rightarrow K'$, $K'\rightarrow K $) 
on the grounds that such scattering transfers large momenta, which makes it a weak process compared to the intravalley scattering. Indeed, the Coulomb interaction in 2D 
scales inversely with the transferred momentum and is therefore small when the momentum transfer is large. We also ignore the interaction $U$ momentum dependence arising from projection onto the conduction band. This is a reasonable approximation because the electron Bloch function momentum dependence is weak in the realistic regime $\epsilon_F\ll D$.





The one-band model, Eq.\eqref{eq:H0}, is well justified in the low-carrier-density regime of interest. 
In this regime, Fermi energy $\epsilon_F$ is much smaller than the BBG bandgap 
by 10 to 100 times
\cite{zhou2022isospin,zhou2021superconductivity, de2021cascade, seiler2021quantum}.  At low carrier density, the dispersion in the conduction band can be approximated as a sum of an isotropic and a trigonal warping term\cite{mccann2013the, jung2014accurate}: 
\be\label{eq:epsilon_KK'}
\epsilon_{K,K'} 
(\vec p)= \sqrt{D^2+\lp \frac{p^2}{2m}\rp ^2} \pm 
\lambda p^3\cos 3\theta_{\vec p} 
,
\ee
where $\vec p$ is the momentum measured from $K$ or $K'$ valley center, $\theta_{\vec p}$ is the azimuthal angle of $\vec p$, 
and $\pm$ corresponds to 
$K$ and $K'$, respectively.
The first term gives an isotropic Fermi sea, 
the trigonal warping term lowers 
the symmetry down to a three-fold rotation symmetry $C_3$. This distortion generates several minima of the curvature on the Fermi surface [see Fig.\ref{fig:FS} c)] --- acting as  hotspots for nesting that drives CDW order. We call these points ``hotspots". 
As we will see, the hotspots are the regions where the CDW/SDW gap tends to open. 
In general, the shape of the Fermi surface can be either convex or partly concave [see Fig.\ref{fig:FS} b) and c)]. For the convex case [see Fig.\ref{fig:FS} b)], the Fermi surface hosts three hotspots $Q_1$, $Q_2$, $Q_3$ associated with each other through rotation symmetry. For the partly concave case [see Fig.\ref{fig:FS} c)], the Fermi surface hosts six hotspots centered at points where curvature vanishes. For simplicity, we will focus on the convex case Fig.\ref{fig:FS} b). A generalization of our analysis to the partly concave case is straightforward and will be discussed below.

 




In the analysis of CDW/SDW, we will need the dispersion expanded around the hotspots. To that end, we define $p_*$ as the radius of curvature at one of the $Q_i$ points ($i=1,2,3$), from now on for conciseness called $Q$. Expanding the electron energy (measured from the Fermi level) around $Q$ gives
\be\label{eq:hotspot_energy}
\epsilon_K 
(Q-K+k)=  
v_F k_\parallel + \frac{k_\perp^2}{2m_*}, \quad m_*=\frac{p_*}{v_F}
,
\ee
valid provided $k\ll |Q-K| $. The dispersion in valley $K'$ is of a similar form, related to that in Eq.\eqref{eq:hotspot_energy}
by time-reversal symmetry. 
The quantity $v_F$ is Fermi velocity, $k_\parallel$ is the momentum parallel to the Fermi velocity and  normal to the Fermi surface, $k_\perp$ is the momentum perpendicular to the Fermi velocity. For simplicity, we will take $v_F$to be constant everywhere on the Fermi surface.

\section{CDW instability at weak coupling}\label{sec:weak-coupling}
To motivate the strong-coupling framework developed in the next section, we start with a conventional weak-coupling analysis of 
the CDW/SDW instability. 
For simplicity, here we only focus on CDW. The analysis and results for SDW are similar and will be discussed later. 

A CDW state with wavevector $2Q$ and amplitude $\Delta$ is described by the
mean-field Hamiltonian 
\begin{align}
H_k &= \lp
\begin{matrix}
	\epsilon_{K}(Q-K+k) 
	& -\Delta 
	 \nonumber \\
	-\Delta 
	&\epsilon_{K'}(-Q-K'+k) 
\end{matrix}
\rp\\
&= \frac{k_\perp^2}{2m_*} \tau_0 - \Delta \tau_1+ v_Fk_\parallel \tau_3, \label{eq:HMF}
\end{align}
where $\tau_{1,2,3}$ are $2\times 2$ Pauli matrices, $\tau_0=1_{2\times 2}$, the CDW order parameter $\Delta$ opens a  gap in small segments on the Fermi surface near each of the hotspots (see Fig.\ref{fig:FS} c)).
The length of each segment is 
\be
k_\Delta = 2\sqrt{2m_*\Delta}.
\ee
The Hamiltonian, Eq. \eqref{eq:HMF}, possesses a $U(1)$ symmetry $\Delta\rightarrow \Delta e^{i\phi}$, resulting in a Goldstone mode corresponding to the sliding motion. As discussed below, this mode is crucial for understanding transport in the CDW phase.



\begin{figure}
	\includegraphics[width=0.4\columnwidth]{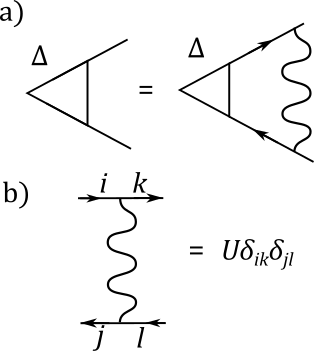}
	\centering
	\caption{a) The self-consistency relation in weak-coupling theory. b) The interaction is written in matrix form.}\label{fig:weak-coupling}
\end{figure}


The onset of CDW instability is predicted by the standard saddle-point self-consistency relation similar to the BCS self-consistency equation for superconductivity:
\be\label{eq:CDW_self-consistency}
\Delta = -\sum_{\epsilon,k}U\Delta \frac{1}{2}{\rm tr} \lb \tau_1
 G(i\epsilon, k)\tau_1 G(i\epsilon, k)\rb   
\ee
where 
$G(i\epsilon, k)$ is a matrix electron Green's function  defined as 
\be\label{eq:G(ie,k)}
G(i\epsilon, k) = 1/\lp i\epsilon-H_k \rp.
\ee 
This equation, in a linearized form, is illustrated diagrammatically in Fig.\ref{fig:weak-coupling}. Linearizing Eq.\eqref{eq:CDW_self-consistency}, and using Eq.\eqref{eq:hotspot_energy} and Eq.\eqref{eq:HMF}, we obtain the following linearized self-consistency relation
\be\label{eq:self_consistency_weak_coupling}
\Delta = -\sum_{\epsilon,k} \frac{U\Delta}{z^2-v_F^2k_\parallel^2},
\ee
where $z = i\epsilon-k_\perp^2/2m_*$, summation over $k$ is carried out over the vicinity of the hotspots $Q_i$ vectors.

To make a comparison with the isospin polarization instability more direct, 
we rewrite this equation in terms of the CDW polarization function $\Pi_{2Q}$. this quantity is nothing but the charge polarization function evaluated at momentum $2Q$. Taking the latter in the free-electron approximation gives
\be\label{eq:Pi_2Q}
1+U\Pi_{2Q} = 0,\quad \Pi_{2Q} =  \frac{1}{2}\sum_{\epsilon,k} {\rm tr} \lp \tau_1
 G^{(0)}_{\epsilon, k}\tau_1 G^{(0)}_{\epsilon, k}\rp
 .
\ee
Here $G_{\epsilon,k}^{(0)}$ is a shorthand form of the free-electron Green's function, Eq.\eqref{eq:G(ie,k)}, 
with $\Delta$ taken to be zero.
This equation is of the same form as the Stoner criterion for valley polarization
\be\label{eq:Pi_v}
1+U\Pi_v=0,\quad \Pi_v =  \frac{1}{2} \sum_{\epsilon,k} {\rm tr} \lp \tau_3 G^{(0)}_{\epsilon, k}\tau_3 G^{(0)}_{\epsilon, k}\rp
\ee 
Since at weak coupling the interaction strength $U$ entering both instabilities is similar in magnitude, 
the competition between CDW instability and the valley polarization instability is decided by the relative strength 
of the polarization functions $\Pi_{2Q}$ and $\Pi_{v}$. We compare these quantities below.

The local flatness of the Fermi surface near the hotspots facilitates nesting and enhances CDW instability. However, while the CDW polarization function $\Pi_{2Q}$ is enhanced, the valley polarization function $\Pi_v$, which is proportional to the total density of states, is mostly insensitive to the curvature of the Fermi surface at the hotspots. As a result, at $T=0$, one expects CDW instability to be stronger and occur before 
the valley polarization instability. 

Yet, the measurement in BBG \cite{zhou2022isospin} does not find isospin-polarized phase being outmatched 
	by CDW phase in most part of the phase diagram. This can be attributed to the critical temperature $T_c$ of the CDW phase being lower than the base temperature of the measurements. Indeed, it is reasonable to expect that $T_c$ values for a weak-coupling CDW are low, since the Fermi surface in BBG does not exhibit a perfect nesting --- the CDW only opens a gap in a small segment on the Fermi surface near the hotspots. As a result, only a small fraction of carriers undergoes condensation, giving a relatively small condensation energy and, therefore, a lower critical temperature $T_c$ as compared to the valley-polarized state in which a large fraction of electrons gains condensation energy.

However, careful examination of the observed phase diagram indicates that the situation is not all that simple. 
	In particular, one question that the weak-coupling scenario does not address is why a CDW phase is seen at the onset of isospin order in the experiment\cite{zhou2022isospin}. In other words, 
why the $T_c$ of the CDW state is enhanced at the isospin ordering phase transition. To illustrate the disagreement between the weak-coupling scenario and measurement, in Fig.~\ref{fig:FS} d), we show a schematic phase diagram, where the cyan dashed line represents the predicted CDW phase boundary in the weak-coupling theory. Here, $T_0$ represents the expected $T_c$ in weak-coupling theory which we assume to lie below the base temperature of the measurement. In comparison, the yellow curve 
	illustrates the behavior of $T_c$ suggested by the measurement\cite{zhou2022isospin}, which lies below the base temperature everywhere except at the onset of isospin order where it is enhanced to a value $T_*$ in the observable range.

This disagreement between the weak-coupling theory and observation calls for a strong-coupling analysis. As we will see in the next section, a strong-coupling analysis predicts an abrupt enhancement of coupling near the onset of isospin order, which pushes the critical temperature of the CDW state there from the low value $T_0$ to a higher value $T_*$. This will resolve the issue pointed out above
and explain the measured phase diagram. 

\section{Strong-coupling framework}\label{sec:CDW_instability}

Here we introduce a strong-coupling framework which accounts for  quantum fluctuations near the isospin phase transition. We assume this transition to be second order (i.e. continuous) or weakly first order, giving rise to soft modes that can assist the CDW/SDW instability. 
To understand the emergence of the CDW/SDW order 
at the onset of isospin polarization order, 
one must determine 
the order type 
which can be either a valley polarization (VP) or intervalley coherence (IVC) order, and in each case identify the soft modes originating from the order parameter fluctuations. 
As discussed below, in both cases soft modes arise in a natural way 
and have distinct impact on the CDW/SDW instability. 
Yet, we do not need to immediately commit to  one of the two possibilities, since 
no matter which order wins, both VP and IVC order parameter fluctuations are likely to be soft near the phase transition.   
Furthermore, measurements\cite{zhou2022isospin} indicate that the phase transition occurs near the van Hove singularity, implying that the system is close to both VP and IVC instabilities.
We will first focus on the soft modes associated with 
a nearby VP instability and then, in Sec.\ref{sec:discussion}, discuss the effects due to a nearby IVC instability. As we will see, proximity to the VP instability 
enhances the CDW/SDW instability whereas the IVC instability suppresses the instability.


Near valley polarization transition, valley-polarizing modes are softened, giving rise to a significant renormalization of the interaction\cite{dong2022spin}. 
To see this, consider the renormalization of the bare intervalley Coulomb interaction. There are two types of renormalization --- screening and vertex correction. Accounting for both of them, we find in Ref.\onlinecite{dong2022spin} and re-derive in Supplement\cite{SM} that the effective interaction $V$ between two electrons in two valleys (one is in valley $K$, the other is in $K'$) 
takes the following form at the critical point:
\be
V(i\nu,q) = \frac{U/N}{\frac{\kappa|\nu|}{2v_F q}+l_0^2 q^2} 
\label{eq:H_eff1}
\ee
Here, $q$ is the momentum transfer, and the quantity $\kappa$ is defined as $\kappa = \frac{p_*+p_*'}{p_0}\sim \kappa_*+1$, where $p_*'$ represents the radius of curvature at $Q'$, which is opposite to $Q$ on the Fermi surface in valley K. The quantity $p_0$ is the average radius of curvature on the whole Fermi sea, $\kappa_*$ is defined as $\kappa_*=p_*/p_0$. We have assumed $p_*'\sim p_0$. The quantity $l_0$ depends on the shape of the Fermi surface. For simplicity, we assume $l_0 \sim k_F$ throughout this paper. The integer $N$ represents the number of isospin species. Here and below, we focus on the unpolarized phase, so $N=4$ (spin $\uparrow$/$\downarrow$, and valley $K$/$K'$). This problem is closely related to a problem tackled by Altshuler, Ioffe, and Millis\cite{Altshuler1994} who showed that in a Fermi sea coupled to gauge field fluctuations, the $2k_F$ vertex is logarithmically enhanced. In Ref.\cite{Altshuler1994}, the transverse gauge propagator takes the same form as Eq. \ref{eq:H_eff1}. Therefore, many of the results of Ref.\cite{Altshuler1994} are directly applicable here, except that in our case the absence of a singular self-energy modifies the answer, as will be seen below.

\begin{figure}
\includegraphics[width=1.0\columnwidth]{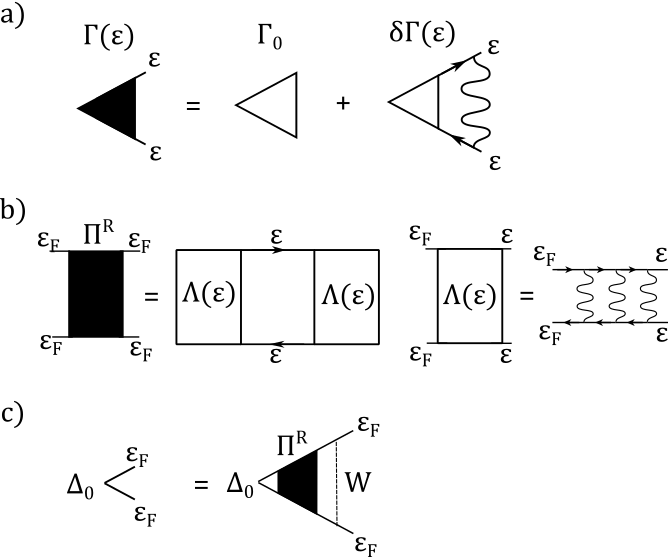}
	\centering
	\caption{A diagrammatic derivation of the self-consistency equation in strong-coupling theory.  a) The renormalization by soft-mode mediated interaction $V$ gives a power-law correction to the CDW vertex $\Gamma(\epsilon)$ for $\Delta \ll \epsilon \ll \epsilon_F$ see Eq.\eqref{eq:RG_power-law}. b) the vertex correction can be absorbed into polarization ladder $\Pi^R_{2Q} $. See Eq.\eqref{eq:PiR}. c) The $T_c$ of CDW/SDW
	can be obtained using the renormalized CDW/SDW polarization function $\Pi^{R}$. See Eq.\eqref{eq:self_consistency_3}.
	 }\label{fig:vertex_correction}
\end{figure}

Due to the frequency dependence of $V(i\nu,q)$, the CDW/SDW vertex function $\Gamma(\epsilon)$ acquires a singular frequency dependence. To see this, we consider the first-order vertex correction shown in Fig.\ref{fig:vertex_correction} a). In this diagram, the frequencies carried by external legs are $\sim \epsilon$, whereas the internal frequencies are considerably higher. 
This first-order vertex correction, evaluated in the Supplement \cite{SM}, is given by
\be\label{eq:1st-order_vertex_correction}
\delta\Gamma(\epsilon) = \frac{Um_*}{\pi N \kappa}\ln\lp \frac{\epsilon_F}{\epsilon}\rp \Gamma_0.
\ee 
where the high-energy cutoff $\epsilon_F$ is the Fermi energy, $\Gamma_0$ is the bare CDW/SDW vertex at the high energy scale $\epsilon_F$. The log divergence 
suggests treating higher-order corrections to the vertex function 
by a renormalization group (RG) approach. Namely, the RG flow of the renormalized vertex $\Gamma^R$ in the regime of 
$\epsilon \ll  \epsilon_F$ is of the form
\be\label{eq:RG}
\frac{d\Gamma^R(\epsilon)}{d\ln{\epsilon}} = -\eta \Gamma^R(\epsilon), \quad \eta = \frac{Um_*}{\pi N \kappa}.
\ee
This gives rise to an infrared power-law divergence: 
\be\label{eq:RG_power-law}
\Gamma^R(\epsilon) = \Lambda(\epsilon) \Gamma_0,\quad \Lambda(\epsilon) = \lp \frac{\epsilon_F}{\epsilon}\rp^\eta.
\ee
This divergence and its implications for CDW order was first discussed in Ref.\cite{Altshuler1994}. 
%

The power-law divergence of the vertex function gives rise to a singular 
CDW/SDW susceptibility. The renormalized CDW susceptibility $\Pi_{2Q}^R$ is obtained by absorbing the vertex correction into the bare CDW susceptibility (see Fig.\ref{fig:vertex_correction} b)). As the energy on ladder in Fig.\ref{fig:vertex_correction} b) flows from high-energy to low-energy and then back to high-energy, $\Pi_{2Q}^R$ is enhanced by a factor of $\Lambda^2(\epsilon)$, 
\be\label{eq:PiR}
\Pi^R_{2Q} = \sum_{\epsilon,k} \frac{\Lambda^2(\epsilon) }{\lp i\epsilon-\frac{k_\perp^2}{2m_*}\rp^2-v_F^2k_\parallel^2},
\ee
which is formally similar to the renormalized CDW susceptibility considered in the problem of fermions coupled to a gauge field [see Ref.\cite{Altshuler1994}, where the renormalized susceptibility is denoted $\Pi(\omega,q)$].  A difference between our results  and those in Ref.\cite{Altshuler1994} is that the self-energy that plays a major role in Ref.\cite{Altshuler1994} does not exist in our problem. This is because the singular interaction here is an intervalley coupling which does not contribute to self-energy. We will discuss this in detail shortly.

We stress that the changes in the interaction due to screening effects and vertex corrections are negligible at a frequency scale of $\epsilon_F$. Therefore, the interaction vertex at $\epsilon_F$ is not the collective-mode-mediated interaction Eq.\eqref{eq:H_eff1}, but a bare interaction denoted as $W_c$ and $W_s$ for CDW and SDW channels respectively. These couplings are represented by dashed line in Fig.\ref{fig:vertex_correction}c), whose strength we will discuss in Sec.\ref{sec:CDW and SDW}. 
Accounting for the ladder diagram in Fig.\ref{fig:vertex_correction}c), we obtain a linearized self-consistency equation in a strong-coupling framework:
\be\label{eq:self_consistency_3}
1+ W_{c/s}\Pi_{2Q}^{R} = 0
\ee
This is essentially the same as in Eq.\eqref{eq:self_consistency_weak_coupling} up to extra vertex correction factors in $\Pi_{2Q}^R$. By power counting, we find the left-hand side is proportional to $T^{1/2-2\eta}$. Therefore, we conclude that either CDW or SDW instability occurs at a significantly enhanced $T_c$ when dimensionless coupling $\eta>1/4$.
The resulting CDW/SDW phase boundary is sketched in Fig.\ref{fig:FS} d) [yellow curve].

The condition $\eta>1/4$ is achievable at the isospin phase transition in BBG. To see this, we plug the Stoner threshold $\frac{U p_0}{2\pi v_F}=1$ into the expression of $\eta$, and find $\eta = \frac{2\kappa_*}{N\kappa}$. Using $N=4$, $\kappa\sim \kappa_*+1$ (see text after Eq.\eqref{eq:H_eff1}) and $\kappa_*\gg1$, we obtain $\eta\sim \frac{1}{2}$. We note parenthetically that the trigonal warping term in the electron Hamiltonian, which breaks the rotation symmetry into a three-fold rotational symmetry, is crucial for CDW. If there was no trigonal warping, then we would find $p_*'=p_*$, so $\kappa=2\kappa_*$ and $\eta=\frac{1}{4}$, which gives a marginal (logarithmic) divergence.

We emphasize that the self-energy correction is ignored in our analysis. This is justified because the self-energy is governed by the intravalley interaction, which does not take the same singular form as the intervalley interaction Eq.\eqref{eq:H_eff1}. Although the same process that generates intervalley interaction still contributes to the intravalley interaction, it is no longer the only leading process. There are those exchange processes that are allowed by the intravalley scattering, which cancel the divergent contribution. This behavior is different from that discussed in Ref.\onlinecite{Altshuler1994} where the self-energy correction due to gauge fluctuations is singular and leads to a reduction of the power-law singularity of the response function. The value of their exponent is $2/3-2\eta$ as opposed to our $1/2-2\eta$ which requires a larger $\eta$ value greater than 1/3 for a divergent $2k_F$ response.

\section{Competition between CDW and SDW orders}\label{sec:CDW and SDW}

The CDW and SDW instabilities are nondegenerate due to the difference in the bare interaction in these two channels $W_c$ and $W_s$.
Indeed, both intervalley Coulomb-induced intervalley scattering $U_{\rm{inter}}$ and phonon-induced intervalley scattering and $U_{\rm ph}$ differentiate $W_c$ and $W_s$.
These two interactions can be written as follows:
\be
H' = \frac{1}{2}\lp U_{\rm{inter}}+U_{\rm{ph}}\rp \psi^\dagger_{K}\psi^\dagger_{K'}\psi_K\psi_{K'}
\ee
where the valley-exchanging electron-electron interaction $U_{\rm{inter}}$ and the phonon mediated interaction\cite{chou2022acoustic} $U_{\rm ph}$ are given by 
\begin{align}
U_{\rm{inter}} &= \frac{e^2}{2\pi K \epsilon_0}\sim 100 \text{ meV nm}^2, \quad\epsilon_0 = 5,\\
U_{\rm ph} &\sim -100 \text{ meV nm}^2
\end{align}
Here, $U_{ph}$ has a negative sign because phonon mediates an attraction. The interaction vertex for CDW instability is now corrected by these two interactions as follows:
\be\label{eq:W_c}
W_{\rm{c}} = U - U_{\rm{inter}} - U_{\rm{ph}}.
\ee 
Here, the extra minus signs are due to enclosing an extra fermionic loop. In comparison, the interaction vertex for SDW is not changed since these two interactions preserve spins 
\be\label{eq:W_s}
W_{s} = U.
\ee
As a result, the competition between CDW and SDW depends on the competition between Coulomb-induced and phonon-induced intervalley scattering. If the former is stronger, then SDW wins, otherwise, the CDW wins.

\section{Observables}\label{sec:Drude_weight}
\begin{figure}
\includegraphics[width=0.8\columnwidth]{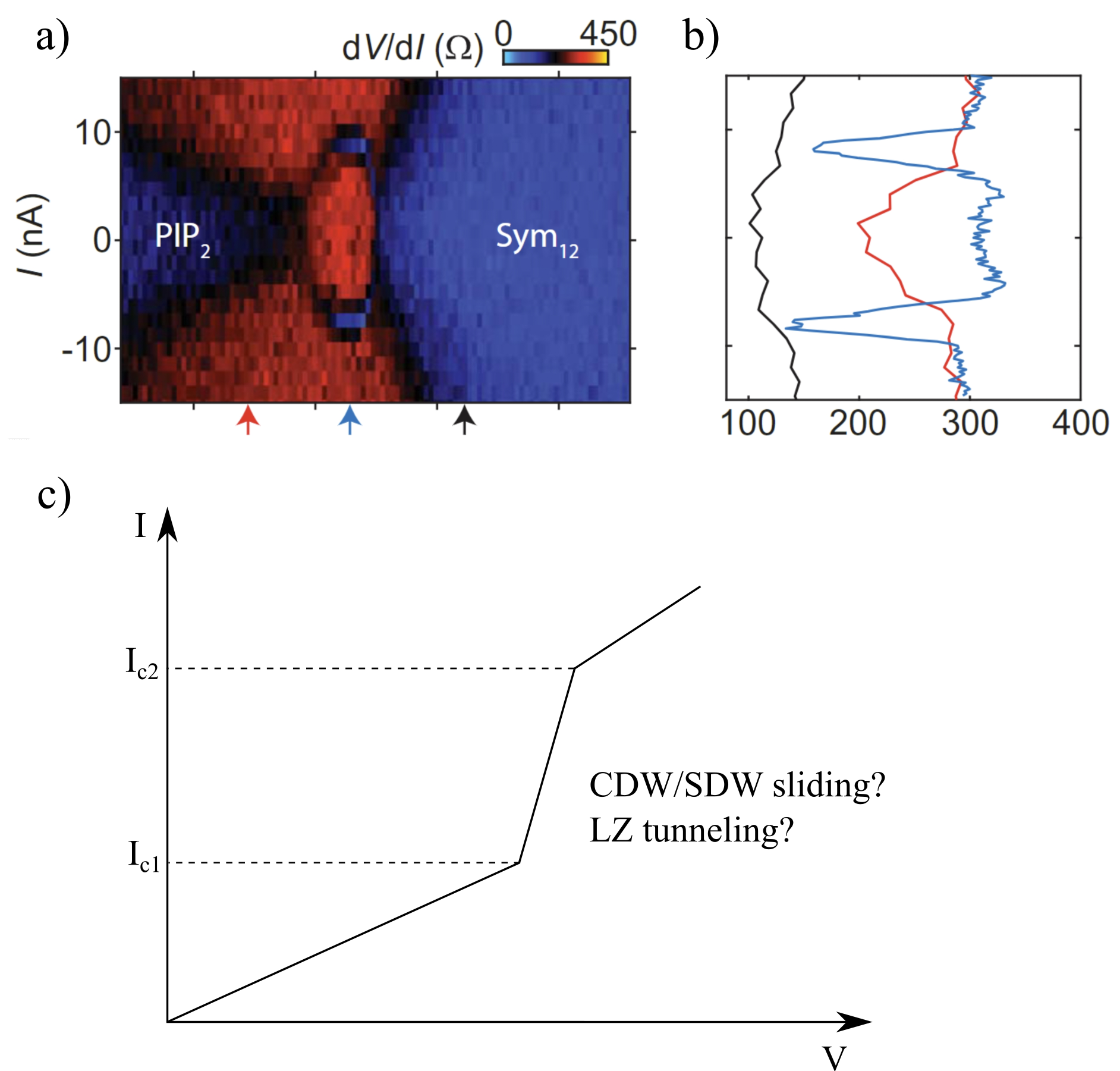}
\centering
\caption{Nonlinear $I$-$V$ dependence extracted from experiment\cite{zhou2022isospin}: a) $dV/dI$ measured at varying carrier density; b) individual line cuts at the three densities indicated by arrows in panel a).  c) Schematic dependence 
$I$ vs. $V$ inferred from b), illustrating an abrupt drop in $dV/dI$ in a finite interval of currents.This behavior agrees with the predictions from two possible scenarios within the CDW framework (see text). In the sliding current scenario, $I_{c1}$ arises from the activation of sliding due to depinning, $I_{c2}$ arises from the CDW breakdown. In the Landau-Zener scenario, the differential resistivity is expected to drop and recover in a narrow regime of current [see Eq.\eqref{eq:Landau-Zener}]. The only quantity that needs to be compared with the experiment is the electric field $E_c$ at which the abrupt change of resistivity occurs. } \label{fig:I-V}
\end{figure}

Experiment\cite{zhou2022isospin} reports on a non-ohmic behavior of conductivity observed near the onset of the isospin-ordered phase, which is reproduced in Fig.\ref{fig:I-V} panels a and b. 
As discussed below, there are several ways this behavior can emerge 
in the CDW phase proposed above. 

Before we begin, it is important to clarify that, unlike CDW in 1D, the CDW phase here is in a metallic state because CDW only gaps out a small segment on the Fermi surface, with the remaining part of the Fermi surface being ungapped. Therefore, conductivity always has an ohmic component, which is not of interest to us. In the analysis below, we will study the contribution from the collective phase mode due to the gapped segment on the Fermi surface, which gives a nonlinear $dV/dI$ (see Fig.\ref{fig:I-V}). 

On general grounds, CDW phase is expected to feature sliding dynamics. As is well known, this dynamics will give rise to a delta-function contribution to conductivity 
$\sigma(\omega) = A \delta(\omega)$\cite{lee1974conductivity}. In one dimension, the current response mediated by CDW sliding is described by
the Drude weight $A=ne^2/m$ which is independent of $\Delta$. 
For our problem, estimates yield  $A$ that depends on $\Delta$ and becomes small when $\Delta$ decreases.
This unconventional behavior arises due to the smallness of the region in $k$-space where nesting occurs. Namely, we obtain
\be\label{eq:weight}
A= C e^2 \sqrt{m_*\Delta}
\ee
where the prefactor $C$ is an order-one dimensionless numerical factor. 
The $\sqrt{\Delta}$ dependence arises from an estimate of the size of the region at the 
Fermi surface in which nesting occurs. This spectral weight $A$ given in Eq.\eqref{eq:weight} is in general much smaller than the standard result\cite{lee1974conductivity} for CDW in one dimension.  
Details for the derivation of Eq.\eqref{eq:weight} are given in Appendix\ref{sec:sliding_mode_conductivity}. 

The CDW sliding conductivity 
can readily explain the measured nonlinear $dV/dI$. Namely, at a low in-plane DC electric field, the CDW is pinned by defects, so the sliding contribution vanishes. At a threshold current $I_{c1}$, the sliding motion is activated, and the $dV/dI$ drops abruptly. When the current is too high, the CDW gap can be suppressed. This can be seen by considering a momentum-$2K$ interband particle-hole pair excitation energy. Such a pair has energy $2\Delta$ in the absence of current. When a current is applied, the standard Doppler shift argument shows that the energy reduces to $\Delta = \Delta-Kv_d$, where $v_d$ is the drift velocity. Therefore, when the applied current reaches a threshold of $I_{c2} = \frac{ne W\Delta}{K} $ (where $W$ represents the width of the sample), the CDW breaks down, and the conductivity $dV/dI$ restores the ohmic value of the high-temperature state. In summary, this mechanism of sliding CDW predicts a $dV/dI$ that rises abruptly at $I_{c1}$ and goes to background value at $I_{c2}$. This behavior qualitatively agrees with measured $dV/dI$. 

For a more quantitative comparison, we extract realistic parameters from experiments. From the measured $T_c$, we estimate the CDW gap as $\Delta=0.01\rm{meV}$. Use realistic parameters $W=1\rm{\mu m}$, $n=0.5\times 10^{12} \rm{cm}^{-2}$, we find $I_{c2}\sim 10 \rm{nA}$, which agrees with the measurement.  Meanwhile, we can also calculate the sliding current, which should be compared with the measured value of $I_{c2}-I_{c1}$. We find the ratio between sliding current and normal current is given by $\sqrt{\frac{\Delta}{\epsilon_F}}\sim 0.1$, predicting a sliding current of $\sim 1\rm{nA}$ when normal current is $10\rm{nA}$. The sliding current extracted from the experiment is $I_{c2}-I_{c1}\approx 5\rm{nA}$ at the same total current, which is a bit larger but still comparable to the expected value.

However, withing the framework of CDW order, there a mechanism that can explain the nonlinear $I(V)$ without invoking CDW sliding dynamics. 
This alternative mechanism 
relies the Landau-Zener (LZ) tunneling of electrons through the CDW gap. Namely, since the CDW gap is quite small,  the electrons can tunnel across the CDW gap when the applied electric field becomes sufficiently strong. The LZ tunneling probability for an electron is\cite{landau1932theorie,zener1932non}
\be
P_{\rm LZ} = \exp \lp -\frac{2\pi \Delta^2}{eEv_F\hbar}\rp \label{eq:Landau-Zener}
\ee
Therefore, we expect activation of the carriers on the gapped segment of the band for fields above the critical value 
\be\label{eq:E_c}
E_c = \frac{\Delta^2}{ev_F\hbar}\sim \frac{\Delta^2 k_F}{e\epsilon_F}.
\ee
The activated carrier density increases abruptly when $E$ approaches $E_c$,
leading to an increase in the conductivity $I/V$. 
Conductivity quickly saturates when $P_{\rm LZ} \rightarrow 1$. This behavior predicts an abrupt step in $I-V$, or equivalently, a narrow dip in the differential resistivity $dV/dI$, resembling the observed dependence (see Fig.\ref{fig:I-V}). 

To assess applicability of this scenario, 
we estimate the critical field $E_c$  and compare to the measured value. Starting with Eq.\eqref{eq:Landau-Zener}, we use the measured $T_c$ value as a crude estimate for the 
CDW gap: $\Delta\sim 0.01\rm{meV}$. 
The band dispersion and Fermi surface calculated numerically\cite{zhou2022isospin} give values $\epsilon_F\sim 1\rm{meV}$ and $k_F\sim 10^{6} \rm{cm}^{-1}$.
 Plugging these into Eq.\eqref{eq:E_c} gives $E_c = 10 \rm{mV/mm}$, whereas the measured value of $E_c$ extracted from the $dV/dI$ measurement\cite{zhou2022isospin} is $3\rm{mV/mm}$. The calculated and measured $E_c$ are of the same order of magnitude. Their difference can be attributed to the uncertainty in our estimate of $\Delta$.

Furthermore, beyond the nonlinear effects in transport, 
the CDW/SDW order can be probed by measuring narrow-band noise under an applied current\cite{gruner1981nonlinear,gruner1981observation,bardeen1982current,nomura1989narrow}. The narrow-band noise frequency $\nu$ is given by
	\be \label{eq:narrow-band noise}
	\nu = v_d/\lambda,
	\ee 
	where $v_d$ is the drift velocity and $\lambda$ is the CDW wavelength.
	Within the CDW sliding scenario, the current above which the resistivity starts to drop can be interpreted as sliding current. The value of this current $I\approx 5\rm{nA}$ 
	allows us to extract the value of  $v_d$ 
	through the relation $v_d=I/neW$, where $W$ is the width of the current-carrying region and $n$ is the carrier density. In Ref.\cite{zhou2022isospin} $W=1\rm{\mu m}$, $n = 0.5\times 10^{12} \rm{cm}^{-2}$. These values yield $v_d= 6 \rm{m}/\rm{s}$. 

However, 
identifying the right value of the modulation wavelength $\lambda$ is somewhat subtle.  The physical meaning of $\lambda$ is the spatial periodicity seen by carriers. Therefore, nominally, its value is nothing but the CDW/SDW wavelength $\lambda= \pi/K\sim 5 \rm{\AA}$, which is on the order of carbon atom spacing. Plugging this into Eq.\eqref{eq:narrow-band noise} yields a high narrow-band noise frequency of $\nu\sim 15\rm{GHz}$. 
Yet, in addition to this frequency, the CDW sliding can also generate 
a lower frequency. This is because in the presence of a IVC order (long-range or short-range)
 the carriers can also sense the long-wavelength modulation of periodicity which is comparable to the Fermi wavelength $\lambda\sim \lambda_F \sim 30\rm{nm}$ (see Fig.\ref{fig:FS}). 
 This would give a lower frequency on the order of $\nu\sim 250\,\rm{MHz}$ which is comfortably low for detection. 
 
 \addLL{We note parenthetically that a short-range IVC order is sufficient for generating this narrow-band noise frequency so long as the correlation length of the IVC order exceeds the periodicity of $30 \rm{nm}$. Such a correlation length is achievable since the system is near a van Hove singularity (vHS). The large density of states at vHS leads to a proximity to the IVC instability as discussed in Sec.\ref{sec:CDW_instability}, resulting in a large IVC correlation length.}



\section{CDW instability suppression in proximity to IVC 
order}\label{sec:discussion}
The analysis in Secs.\ref{sec:CDW_instability} and \ref{sec:CDW and SDW} is based on the assumption that the region of CDW/SDW onset is close to a VP instability. However, as discussed in Sec.\ref{sec:CDW_instability}, 
 a competing intervalley coherence (IVC) instability near the region of interest may exist. Does a similar strong-coupling CDW/SDW scenario work near an IVC instability? 
It can be seen that the answer to this question is negative. As a matter of fact, the CDW/SDW instability is suppressed 
near IVC instability because the soft-mode interaction arising from this instability\cite{chatterjee2022inter,dong2023signatures} has an opposite sign compared to that of the VP instability. 
\addLL{Indeed,} each IVC soft-mode interaction line \addLL{when plugged} into the CDW vertex correction 
introduces an extra fermionic loop compared to the VP soft-mode interaction (Eq.\ref{eq:H_eff1}), giving an extra minus sign. As a result, the IVC soft mode tends to suppress the CDW/SDW instability.

It is instructive to compare this conclusion to \addQ{the scenario 
\addLL{of a proximal}	superconducting (SC) phase} developed in Ref.\cite{dong2023signatures}. As a reminder, this SC phase 
	is observed at almost the same carrier density and transverse field as the resistive phase, 
except that the SC phase occurs above a finite in-plane magnetic field $B_\parallel$ whereas the resistive phase in question occurs at $B_\parallel=0$\cite{zhou2022isospin}.
	In Ref.\cite{dong2023signatures} it was shown that the measured phase diagram points to a scenario where this superconductivity is mediated by soft modes near an IVC instability. However, 
	\addLL{here we interpret the resistive phase} 
	as a CDW/SDW phase that arises near the onset of VP. At a first glance it might seem that these two 
	\addLL{approaches contradict} 
	each other, as they resort to different types of isospin instabilities. However, the two scenarios 
	\addLL{are not mutually exclusive} once we recognize that the phase transition 
	through which SC and CDW/SDW order occurs is close to the van Hove singularity\cite{zhou2022isospin}. In this situation, the divergent density of states predicts that \textit{both} IVC and VP instabilities are strong\addLL{, however which of them actually wins is unimportant}. 
	
Furthermore, the competition between SC and CDW/SDW and its dependence on an in-plane magnetic field $B_\parallel$ can also be understood within \addQ{ a scenario of coexisting and competing IVC and VP instabilities}
	. Namely, it is shown in Ref.\cite{dong2021superconductivity} that $B_\parallel$ tends to suppress the low-frequency singularity of VP-mode interaction, which is the interaction used by CDW/SDW. This suppression arises due to the screening effect between two spin species that are nondegenerate under $B_\parallel$. In comparison, we do not expect the low-frequency singularity of the IVC-soft-mode interaction to be suppressed since this interaction is not related to screening. This distinction between the two soft-mode interactions predicts a suppression of CDW/SDW upon applying $B_\parallel$. This prediction agrees with the behavior reported in Ref.\cite{zhou2022isospin}.
	

\section{Conductivity due to sliding mode}\label{sec:sliding_mode_conductivity}
In order to compare with the measurement, we calculate the sliding mode contribution to conductivity $\delta \sigma$.  Although the exact value of it depends on relaxation time, the ratio between $\delta \sigma$ and the normal ohmic conductivity $\sigma_0$ is nothing else but the ratio between the Drude weights of these two contributions, which is independent of the relaxation time. In this section, we will obtain this ratio, which is used in main text when comparing with measurement.

As a reminder, the conductivity of CDW/SDW consists of two parts: the single-particle contribution and the collective mode contribution. For CDW/SDW in 2D, the CDW/SDW gap only gaps out a segment on the Fermi surface, therefore the single particle contribution consists of two parts: one arises from the gapped segment, which gives two wings outside the CDW/SDW gap $\Delta$, the other comes from the gapless segments, which gives a Drude peak. Similar to the one dimensional case, the collective mode contribution to 2D CDW/SDW also gives a correction to the Drude weight. As we will see below, the quantity that can be directly compared with the measurement is the collective mode contribution to Drude weight. Therefore, we will focus on this contribution below. 

\begin{figure}
\includegraphics[width=0.9\columnwidth]{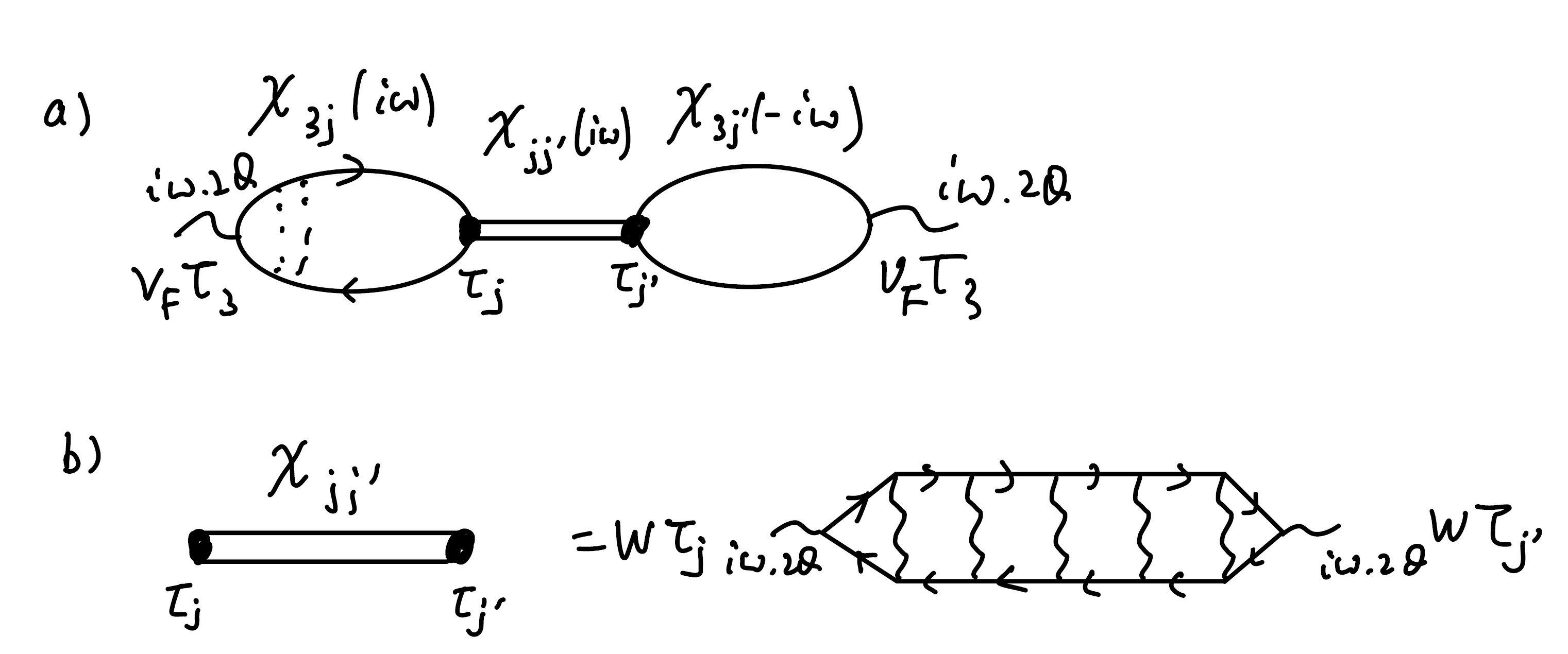}
	\centering
	\caption{a) The collective mode's contribution to current-current susceptibility b) the collective mode propagator. See Eq.\eqref{eq:D}}\label{fig:collective_mode}
\end{figure}

To calculate conductivity, we \addLL{consider} the AC susceptibility 
\be\label{eq:chi_def}
\chi(i\omega)=\frac{\delta j}{\delta A},
\ee
\addLL{where $A$ is the electromagnetic vector potential}. The quantity $\chi(i\omega)$ is related,  up to an $i\omega$ factor, to AC conductivity $\sigma(i\omega)$. Naturally, here and below we focus on the longitudinal susceptibility in which the vectors $\vec j$ and $\vec A$ are parallel. 
The collective mode contribution to $\chi$ is diagrammatically expressed in Fig.~\ref{fig:collective_mode} a), where each line represents the electron Green's function 
\begin{align}\label{eq:electron_green's_function_appendix}
&G(i\epsilon,k) = \frac{z-\Delta \tau_1 + v_Fk_\parallel \tau_3}{z^2-E^2} ,\\
&z = i\epsilon -\frac{k_\perp^2}{2m_*}, \quad E = \sqrt{v_F^2k_\parallel^2 +\Delta^2}. \label{eq:def_E}
\end{align}
Then the collective mode contribution to $\chi$ defined in Eq.\eqref{eq:chi_def} can be written as:
\begin{align}
\delta\chi(i\omega)  &= -v_F^2  \sum_{j,j'=0,1,2,3} \sum_\epsilon {\rm tr} 
\lp G_- \tau_3 G_+ \tau_j \lp \frac{\epsilon_F}{\epsilon}\rp^{\eta} \rp 
\nonumber \\
& \times (-\chi_{jj'}(i\omega)) \sum_{\epsilon'}{\rm tr} 
\lp G_+ \tau_3 G_- \tau_{j'}\lp \frac{\epsilon_F}{\epsilon'}\rp^{\eta} \rp
\end{align}
Here and below the electron Green's function $G_\pm = G(i\epsilon_\pm) = G(i\epsilon\pm i\frac{\omega}{2}) $. The first minus sign arises from the fermionic loop, the trace notation is defined so that it includes integrals over momentum ${\rm tr} (...)=\sum_{k} {\rm tr} (...)$,
and the collective-mode propagators $\chi_{jj'}$'s are diagrammatically defined as in Fig.\ref{fig:collective_mode} b), and can be written as 
\begin{align}\label{eq:D}
\chi_{jj'}(i\omega) &= -W^2\int d\tau e^{i\omega\tau} \langle T \rho_j(\tau) \rho_{j'}(0) \rangle, \\
\rho_j &= \Psi^{\dagger} \tau_j\Psi , \quad j=0,1,2,3.
\end{align}
we find only $\chi_{22}$ gives a contribution since $\chi_{31},\chi_{30}=0$. In the  
\begin{align}
\label{eq:delta chi}
\delta\chi(i\omega) &= -v_F^2 \chi_{32}(i\omega)   \chi_{22}(i\omega) \chi_{32}(-i\omega) ,\\
\chi_{32}(i\omega) &= \sum_\epsilon {\rm tr} \lp G_- \tau_3 G_+ \tau_2 \lp \frac{\epsilon_F}{\epsilon}\rp^{2\eta} \rp 
\end{align}
\addQ{This expression of $\delta\chi(i\omega)$ is overall similar to the usual response function in CDW\cite{lee1974conductivity}, except that both current vertices in it are renormalized by the interaction at critical point. Therefore, this expression contains extra the power-law renormalization factors.}  Here, the quantity $\chi_{22}(i\omega)$ is nothing but the sliding mode propagator, which is diagrammatically described in Fig.\ref{fig:collective_mode} and is expressed as follows:
\be\label{eq:chi_22_1}
\chi_{22}(i\omega) = \frac{W}{1 + W \Pi_{2}^R(i\omega)},
\ee
where $W$ is the bare interaction defined 
in Eq.\eqref{eq:W_c} and Eq.\eqref{eq:W_s} for CDW and SDW respectively. The quantity $\Pi_{2}^R(i\omega)$ is the intervalley phase susceptibility renormalized by the critical fluctuations, which is defined as follows:
\be
\Pi_{2}^R(i\omega) = \frac{1}{2}\sum_{k,\epsilon}{\rm tr} \lb G_- \tau_2 G_+ \tau_2 \lp \frac{\epsilon_F}{\epsilon}\rp^{2\eta}\rb 
\ee
As a reminder $G_\pm$ is defined as $ G_\pm= G(i\epsilon\pm i\frac{\omega}{2})$. 

\addQ{On general grounds, we expect that the sliding-mode propagator $\chi_{22}(i\omega)$ has a pole at $\omega=0$ 
as the CDW sliding mode is a Goldstone mode. 
In the ladder response framework used to analyze  $\chi_{22}(i\omega)$  the condition for having a pole at $\omega=0$ is that}
$\Pi_{2}^R$ satisfies
\be\label{eq:1+Uchi22}
1+W\Pi_{2}^R(0) = 0.
\ee 
To confirm that this relation holds, we calculate $\Pi_{2}^R(0)$ explicitly:
\begin{align}
\Pi_{2}^R(0) &= \addQ{\frac12}\sum_{\epsilon,k} {\rm tr} (G \tau_2 G \tau_2 \lp \frac{\epsilon_F}{\epsilon}\rp^{2\eta}) \nonumber\\
&=  \sum_{\epsilon,k}\frac{ \lp i\epsilon-\frac{k_\perp^2}{2m_*}\rp^2-v_F^2k_\parallel^2-\Delta^2 }{\lb \lp i\epsilon-\frac{k_\perp^2}{2m_*}\rp^2-v_F^2k_\parallel^2-\Delta^2 \rb^2}\lp \frac{\epsilon_F}{\epsilon}\rp^{2\eta} \nonumber\\
&= \sum_{\epsilon,k}\frac{ \lp \frac{\epsilon_F}{\epsilon}\rp^{2\eta}}{ \lp i\epsilon-\frac{k_\perp^2}{2m_*}\rp^2-v_F^2k_\parallel^2-\Delta^2 }\label{eq:Pi_2^R}
\end{align}
Importantly, the quantity $\Delta$ is itself determined by the CDW/SDW self-consistency equation, which takes the form: 
\be \label{eq:self_consistency_in_CDW}
\sum_{\epsilon,k} -\frac{W\lp \frac{\epsilon_F}{\epsilon}\rp^{2\eta} }{\lp i\epsilon-\frac{k_\perp^2}{2m_*}\rp^2-v_F^2k_\parallel^2-\Delta^2} = 1
\ee
\addQ{Comparing Eq.\eqref{eq:Pi_2^R} and Eq.\eqref{eq:self_consistency_in_CDW} we confirm the Goldstone mode condition given in Eq.\eqref{eq:1+Uchi22}}.

To summarize this part of the analysis, 
Eq.\eqref{eq:1+Uchi22} is not a coincidence because  it follows from a physical argument: The sliding mode, whose propagator is $\chi_{22}(i\omega)$, is the Goldstone mode arising from the system's $U(1)$ symmetry $\Delta\rightarrow \Delta e^{i\phi}$ (see Hamiltonian Eq.\eqref{eq:HMF}). Therefore, its frequency vanishes, indicating that the pole of $\chi_{22}(i\omega)$ at $\omega=0$. In other words, the denominator of $\chi_{22}(i\omega)$, which equals to $1+W\Pi_{22}(0)$, must vanish at $\omega=0$. 

Given these observations, the expression for $\chi_{22}(i\omega)$ in Eq.\eqref{eq:chi_22_1} can be put in the form that makes the Goldstone mode pole more apparent.
Using the relation Eq.\eqref{eq:1+Uchi22}, we can write $\chi_{22}(i\omega)$ as
\be\label{eq:chi22}
\chi_{22}(i\omega) = \frac{1}{\Pi_{2}^R(i\omega) -  \Pi_{2}^R(0)}.
\ee 
\addQ{This expression does not include the interaction strength $W$, which enables us to evaluate $\chi_{22}(i\omega)$ directly from $\Pi^R(i\omega)$. Both the $\omega=0$ pole and the residue at this pole can be obtained by analyzing the behavior of $\Pi^R(i\omega)$ at small $\omega$.  }
We calculate the quantities $\chi_{32}(i\omega)$ and $\chi_{22}(i\omega)$, which is presented in detail in next section. 
Plugging the results into Eq.\eqref{eq:delta chi}, we find that the collective mode contribution to dynamical current-current susceptibility is 
\be\label{eq:delta chi result}
\delta\chi(i\omega) = - C v_F\sqrt{m_*}\Delta^{\frac{1}{2}} 
\ee
where $C \sim O(1)$ is an order-one factor. This indicates an extra Drude weight, which gives an extra contribution to zero-frequency conductivity:
\be\label{eq:sliding conductivity_appendix}
\delta\sigma(\omega) = -\frac{e^2v_F\sqrt{m_*}\Delta^{\frac{1}{2}} }{i\omega} 
\ee
\addQ{In next section, we will detail the derivation of Eq.\eqref{eq:delta chi result}. }

\section{
The collective mode contribution to the AC susceptibility} 
\addQ{In last section, we have outlined the steps to obtain the collective mode's contribution to AC susceptibility $\delta\chi(i\omega)$ Eq.\eqref{eq:sliding conductivity_appendix}. In this section, we detail the derivation of this result explicitly. }
From Eq.\eqref{eq:delta chi} we see that to calculate $\delta\chi(i\omega)$ we need to evaluate $\chi_{32}(i\omega)$ and $\chi_{22}(i\omega)$. Below, we evaluate these two quantities.

First, the quantity $\chi_{32}(i\omega)$, which is represented by a bubble at each end of the diagram in Fig.\ref{fig:collective_mode} \addQ{a)}, is given by
\begin{align}
\chi_{32}(i\omega) & = \int \frac{d\epsilon}{2\pi} \sum_k \frac{1}{2}{\rm tr} (G(i\epsilon_-) \tau_3 G(i\epsilon_+) \tau_2)\lp \frac{\epsilon_F}{\epsilon}\rp^{\eta} \nonumber \\
&= \int_{-\epsilon_F}^{\epsilon_F} \frac{d\epsilon}{2\pi} \sum_k \frac{iz_- (-\Delta) -iz_+ (-\Delta)}{\lp z_+^2-E^2 \rp \lp z_-^2-E^2 \rp } \lp \frac{\epsilon_F}{\epsilon}\rp^{\eta}\nonumber\\
&= \int_{-\epsilon_F}^{\epsilon_F} \frac{d\epsilon}{2\pi} \sum_k \frac{-\omega\Delta\lp \frac{\epsilon_F}{\epsilon}\rp^{\eta}}{\lp z_+^2-E^2 \rp \lp z_-^2-E^2 \rp }\nonumber
\end{align}
In the first line, $\epsilon_\pm = \epsilon \pm \frac{\omega}{2}$, in the second line we defined $z_\pm = i\epsilon_{\pm} -\frac{k_\perp^2}{2m_*}$. As a reminder here $E$ is \addQ{the quasiparticle energy} defined through $E = \sqrt{v_F^2k_\parallel^2 +\Delta^2}$. To proceed, we keep only the terms leading-order in $\omega$, which gives: 
\begin{align}
\chi_{32}(i\omega) = -\omega\Delta\int\limits_{-\epsilon_F}^{\epsilon_F}\frac{d\epsilon dk_\perp dk_\parallel}{(2\pi)^3} \frac{\lp \frac{\epsilon_F}{\epsilon}\rp^{\eta}}{\lp v_Fk_\parallel -z \rp^2 \lp v_Fk_\parallel +z \rp^2 } 
,
\end{align}
where we suppressed the higher order terms $O(\omega^3)$. 
As a reminder $z$ is defined in the same way as in Eq.\eqref{eq:electron_green's_function_appendix}, namely, $z = i\epsilon -\frac{k_\perp^2}{2m_*}$.

Next, we carry out integral over $k_\parallel$ and $k_\perp$ step by step as follows:
\begin{align} \label{eq:introduce epsilon_perp}
&\chi_{32}(i\omega) = -\frac{\omega\Delta}{v_F}\int_{-\epsilon_F}^{\epsilon_F}\frac{d\epsilon}{2\pi} \frac{dk_\perp}{2\pi}\frac{(-2)i}{ \lp 2z \rp^3 } \lp \frac{\epsilon_F}{\epsilon}\rp^{\eta}
\\
&= \frac{i\omega\Delta}{4v_F}\int_{-\epsilon_F}^{\epsilon_F}\frac{d\epsilon}{2\pi} 
\int_0^{\infty}\frac{d\epsilon_\perp}{2\pi}\frac{\sqrt{2m_*}}{\sqrt{\epsilon_\perp}} \frac{\epsilon_\perp^3 - 3\epsilon^2\epsilon_\perp}{ -\lp\epsilon^2_\perp +\epsilon^2\rp^3 } \lp \frac{\epsilon_F}{\epsilon}\rp^{\eta}.
\nonumber
\end{align}
Here, 
we defined $\epsilon_{\perp} = \frac{k_\perp^2}{2m_*}$. 
\addLL{To carry out integration, we nondimensionalize $\epsilon_\perp=x\epsilon$ and integrate over $x$ as follows:}
\begin{align}
    &-\frac{i\omega\Delta\sqrt{2m_*}}{4v_F}\int_{-\epsilon_F}^{\epsilon_F}\frac{d\epsilon}{2\pi} \int_0^{\infty}\frac{dx}{2\pi} \frac{x^{\frac{5}{2}} - 3x^{\frac{1}{2}}}{ \lp x^2 +1\rp^3 } \frac{|\epsilon|^{\frac{5}{2}}|\epsilon|}{|\epsilon|^6}\lp \frac{\epsilon_F}{\epsilon}\rp^{\eta}\nonumber\\
&=\frac{3}{16\sqrt{2}}\frac{i\omega\Delta\sqrt{2m_*}}{4v_F}\int_{-\epsilon_F}^{\epsilon_F}\frac{d\epsilon}{2\pi} |\epsilon|^{-\frac{5}{2}}\lp \frac{\epsilon_F}{\epsilon}\rp^{\eta}. 
\label{eq:integrals}
\end{align}
To arrive at the expression in the last line [Eq.\eqref{eq:integrals}] we have used the identity 
\be\int_0^{\infty}\frac{dx}{2\pi} \frac{x^{\frac{5}{2}} - 3x^{\frac{1}{2}}}{ \lp x^2 +1\rp^3 } =\frac{3}{64\sqrt{2}} - \frac{15}{64\sqrt{2}} = -\frac{3}{16\sqrt{2}}.
\ee
Note that the factor of $\lp \frac{\epsilon_F}{\epsilon}\rp^{\eta}$ 
needs to be cut off at an energy scale of $\epsilon\sim \Delta$. Therefore, we \addQ{ carry out the integral over $\epsilon$ exactly while approximately cutting off} the integral in Eq.\eqref{eq:integrals} at \addQ{a lower-bound of} $\epsilon\sim \Delta$: 
\addLL{
\begin{align}
\chi_{32}(i\omega)
&= i\omega\frac{3\Delta\sqrt{m_*}}{64\pi v_F } 2\int _\Delta^{\epsilon_F} \lp\frac{\epsilon_F}{|\epsilon|}\rp^\eta  |\epsilon|^{-\frac{5}{2}} d\epsilon\nonumber\\
&= i\omega\frac{3\Delta\sqrt{m_*}}{32v_F}  \frac{1}{\eta+3/2}\lp\frac{\epsilon_F}{\Delta}\rp^\eta  \Delta^{-\frac{3}{2}}.
\end{align}
}

Next, we evaluate the quantity $\chi_{22}(i\omega)$, which is the sliding mode propagator diagrammatically represented by the double line in the middle of Fig.\ref{fig:collective_mode} c). As argued above [see Eq.\eqref{eq:chi22} and accompanying discussion], this quantity is \addLL{the inverse of} 
the difference \be\label{eq:Pi(w)-Pi(0)}
\Pi\addQ{_{2}^R}(i\omega) -  \Pi_{2}^R(\omega\to 0).
\ee
We therefore proceed to evaluate the quantity $\Pi_{22}(i\omega)$, which is given by
%
%
\begin{align}
&\Pi\addQ{_{2}^R}(i\omega) 
=\addQ{\frac{1}{2}}\int \frac{d^2k}{(2\pi)^2}\int\limits^{\epsilon_F}_{-\epsilon_F} \frac{d\epsilon}{2\pi} {\rm tr} (G(i\epsilon_+) \tau_2 G(i\epsilon_-) \tau_2))\lp \frac{\epsilon_F}{\epsilon}\rp^{2\eta} 
\nonumber
\\
&= \addQ{\frac{1}{2}}\int \frac{d^2k}{(2\pi)^2}\int\limits^{\epsilon_F}_{-\epsilon_F} \frac{d\epsilon}{2\pi} \lb  \frac{1}{i\epsilon_+-\frac{k_\perp^2}{2m_*}-E}\frac{1}{i\epsilon_--\frac{k_\perp^2}{2m_*}+E} \right.
\\
&\left. + \lp E\rightarrow -E\rp \rb \lp \frac{\epsilon_F}{\epsilon}\rp^{2\eta}. 
\nonumber
\end{align}
Here $\epsilon_\pm$ is defined as $\epsilon_\pm = \epsilon \pm \frac{\omega}{2}$.
Next, we 
carry out the integral over $\epsilon$. This cannot be done analytically, but we observe that the contribution to the integral over $\epsilon$ predominantly comes from energy $\epsilon\addQ{\sim \Delta}$.  
\addQ{This is because of the following two reasons, a) 
the power-law factor $(\epsilon_F/\epsilon)^\eta$ becomes largest at small energy, b) however, this power-law behavior is valid only for $\epsilon\gtrsim \Delta$ and therefore should be cutoff at an energy of$\epsilon\sim \Delta$. } 
Consequently,
\addQ{as an approximation,} we can replace $\lp \frac{\epsilon_F}{\epsilon}\rp^{\eta}$ with $\lp\frac{\epsilon_F}{\Delta}\rp^\eta $. With this, we can proceed analytically as follows:
\begin{align}
&\Pi\addQ{_{2}^R}(i\omega) = 
\addQ{\frac{1}{2}} \lp\frac{\epsilon_F}{\Delta}\rp^{2\eta}\int \frac{d^2k}{(2\pi)^2} \frac{1}{-i\omega+2E}\lb  f \lp \frac{k_\perp^2}{2m_*}+E\rp \right.
\nonumber\\
& \left. - f \lp \frac{k_\perp^2}{2m_*}-E\rp \rb + \lp E\rightarrow -E\rp  
\nonumber\\
&= \lp\frac{\epsilon_F}{\Delta}\rp^{2\eta}\int \frac{d^2k}{(2\pi)^2} \lb\frac{1}{-i\omega+2E}- \frac{1}{-i\omega-2E}\rb
\nonumber\\
&\times \lb  f \lp \frac{k_\perp^2}{2m_*}+E\rp - f \lp \frac{k_\perp^2}{2m_*}-E\rp \rb 
\end{align}
\addQ{As $E$ is independent of $k_\perp$ (see Eq.\eqref{eq:def_E}), it is convenient to first integrate over $k_\perp$. This integration can be carried out exactly because the last factor simply takes value of $-1$ for $|k_\perp^2/2m|<E$, and vanishes for $|k_\perp^2/2m|>E$. Consequently, the integral over $k_\perp$ yields a factor of $2\sqrt{2m_* E}$, which physically represents the length of the segment on the Fermi surface that is gapped out by CDW}
\begin{align}
&\Pi\addQ{_{2}^R}(i\omega) 
= -\lp\frac{\epsilon_F}{\Delta}\rp^{2\eta}\int \frac{dk_\parallel}{2\pi} \frac{-\addQ{2}E}{-\frac{\omega^2}{4}-E^2} \frac{2\sqrt{2m_*E}}{2\pi}   
\end{align}
Next, \addQ{to integrate over $k_\parallel$, we define $\epsilon_\parallel = v_F k_\parallel$, and rewrite the expression as an integral over $\epsilon_\parallel$}
\begin{align}
&\Pi\addQ{_{2}^R}(i\omega) 
= -\frac{\sqrt{2m_*}}{
\pi^2v_F}\lp\frac{\epsilon_F}{\Delta}\rp^{2\eta} \int_{\addQ{-\infty}}^{\addQ{\infty}}  \frac{d\epsilon_\parallel \lp\epsilon_\parallel^2+\Delta^2\rp^\frac{3}{4}}{\frac{ \omega^2}{4}+\epsilon_\parallel^2+\Delta^2}   
\label{eq:epsilon_parallel}
\end{align}
Taking the difference $\Pi\addQ{_{2}^R}(i\omega) -  \Pi\addQ{_{2}^R}(0)$
and keeping only the leading-order terms in $\omega$, we find
\begin{align}
\Pi\addQ{_{2}^R}(i\omega) -  \Pi\addQ{_{2}^R}(0) &= \frac{\omega^2}{4} \frac{\sqrt{2m_*}}{
\pi^2v_F}\lp\frac{\epsilon_F}{\Delta}\rp^{2\eta}\int\limits_{\addQ{-\infty}}^{\addQ{ \infty}}  \frac{d\epsilon_\parallel}{\lp \epsilon_\parallel^2+\Delta^2\rp^{\frac{5}{4}}}  \nonumber\\
& \sim  \zeta \frac{\omega^2}{2} \frac{\sqrt{2m_*}}{
\pi^2v_F}\lp\frac{\epsilon_F}{\Delta}\rp^{2\eta} \Delta^{-\frac{3}{2}}  
\end{align}
\addQ{where the dimensionless constant $\zeta$ is defined as $\zeta = \int^{\infty}_0 dx (x^2+1)^{-5/4}$, which arises from the integral over $\epsilon_\parallel$. As $\zeta$ is an order-1 prefactor, we will simply take it to be $1$ in the followings steps.}
Plugging this into Eq.\eqref{eq:chi22}
yields
\be\label{eq:chi_22_result}
\chi_{22} = \frac{\addQ{2}\pi^2v_F \lp\frac{\Delta}{\epsilon_F}\rp^{2\eta}\Delta^{\frac{3}{2}}}{\sqrt{2m_*}}\frac{1}{\omega^2}
\ee 
Finally, using Eq.\eqref{eq:delta chi}, we find that 
the collective mode contribution to dynamical susceptibility is given by
\bea
\delta \chi(i\omega) &&= -v_F^2 \chi_{32}(i\omega)\chi_{22}(i\omega)\chi_{32}(-i\omega)\nonumber\\
&&=-\frac{\addQ{2}\pi^2v_F\Delta^{\frac{3}{2}} }{\sqrt{2m_*}} \lp\frac{3\Delta\sqrt{m_*}}{32}  \Delta^{-\frac{3}{2}} \rp^2 \nonumber\\
&&=-\frac{9\pi^2v_F\sqrt{m_*}\Delta^{\frac{1}{2}} }{512\sqrt{2}} 
\eea
As shown in Eq.\eqref{eq:sliding conductivity_appendix}, this quantity represents the \addQ{contribution to the }spectral weight of the \addQ{AC }conductivity \addQ{arising from} 
CDW sliding mode. As discussed above, this spectral weight is independent of total carrier density since the CDW gap is weak, $\Delta\ll \epsilon_F$. Instead, the spectral weight is determined by the length of the segment on the Fermi surface that is gapped out by CDW order, which makes it proportional to $\Delta^{\frac12}$. 

\section{Summary}

Motivated by recent experimental observations \cite{zhou2022isospin, seiler2023interaction}, 
we considered a density-wave phase (charge or spin) near the isospin phase transition in BBG driven by the critical valley fluctuations. Strong interactions between critical fluctuations and band electrons define a unique strong-coupling CDW phase in which the interactions are characterized by momentum transfer much smaller than $k_F$. We use a renormalization group approach to show that the CDW and SDW susceptibilities have a power-law divergence, indicating that $T_c$ is enhanced compared to CDW/SDW emerging away from criticality. 
The predicted phase diagram is consistent with the transport measurement\cite{zhou2022isospin}, where a peculiar resistive state is observed directly at an onset of isospin order.
%

We use this strong-coupling framework to analyze nonlionear transport. In particular, we explain the anomalous non-monotonic differential conductivity behavior reported in Ref.\onlinecite{zhou2022isospin} and argue that it can be understood in the framework of the conventional CDW/SDW phase mode sliding mechanism. We estimate the sliding conductivity in the strong-coupling framework and arrive at a result which is in reasonable agreement with measurements. We also consider an alternative explanation of the nonlinear transport based on the interband Landau-Zener tunneling across the minigaps induced by the CDW/SDW perturbation, finding that this scenario can also reproduce the observed nonlinear behavior. Together, these results provide a strong indication that the resistive state seen in the experiment is indeed related to CDW/SDW order mediated by quantum-critical interactions.


\bibliography{ref}

	


	


\newpage

\onecolumngrid
\appendix

\section{The intervalley interaction mediated by valley-polarization soft mode}
In this section, we describe the origin of singular intervalley interaction used above 
in Eq.\eqref{eq:H_eff1} to analyze CDW and SDW instabilities. This analysis follows closely the analysis carried out in Ref.\cite{dong2022spin}.

\begin{figure}[h]
	\includegraphics[width=0.5\textwidth]{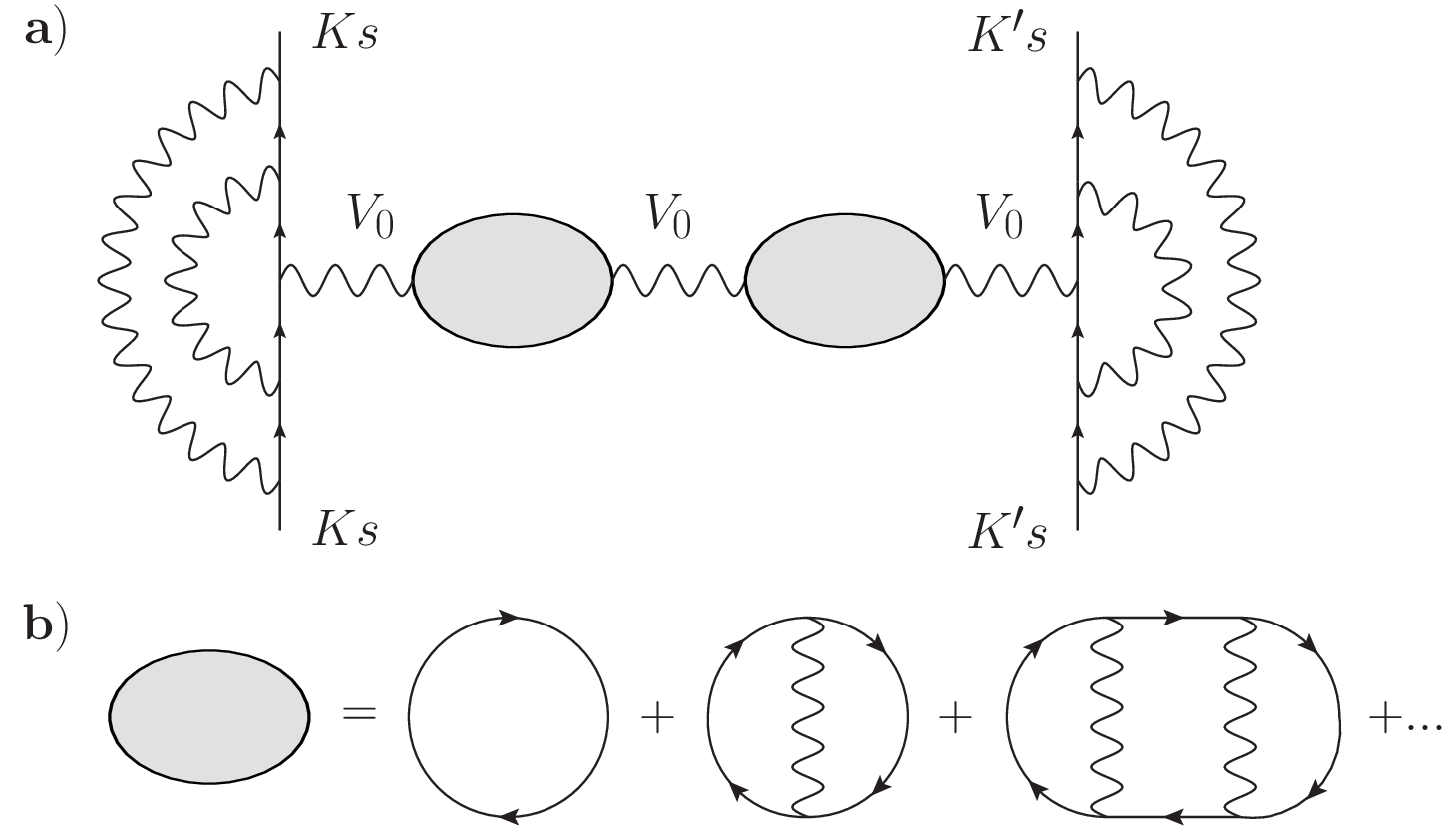}
	\caption{[Figure adapted from Ref.\cite{dong2022spin}] Diagrams describing the effective interaction between electrons in valleys $K$ and $K'$ mediated by quantum-critical modes, Eq.\eqref{eq:sup_V}. Here, the wavy lines represent bare Coulomb interaction $U$. These processes give an enhancement to forward scattering divergent near the valley-polarization instability (see discussion in Ref.\cite{dong2022spin}). (b)The diagrammatic representation of the irreducible part of the charge susceptibility (the shaded ellipse), summed over $s=\uparrow,\downarrow$.}
	\label{fig:soft-mode interaction}
\end{figure}

The intervalley interaction is dominated by the intravalley scattering processes, where an electron in valley $K$ is scattered to valley $K$, and the other electron in valley $K'$ is scattered to valley $K'$. Here we can safely neglect the valley-exchange interaction that transfers a large momentum $K$ since the Coulomb interaction is proportional to the inverse of momentum transfer. At the isospin-polarizing phase transition, such intervalley scattering is renormalized by two effects: the vertex corrections and screening. The renormalized interaction is diagrammatically expressed in \addQ{Fig.\ref{fig:soft-mode interaction}}. Here, the wavy lines represent bare Coulomb interaction $U$.

Below we calculate the renormalized intervalley interaction. First, the effect of each vertex correction can be described using a factor $\gamma$:
\be
\gamma(i\nu,q) = \frac{1}{1+ U \Pi_0(i\nu,q)}
\ee
Here, $\Pi_0$ is the bare polarization function of one spin species, $\Pi_0(i\nu, q) = \sum_{\epsilon,k} G(i\epsilon+i\nu,k+q)G(i\epsilon,k)$. In our setting, spin up and spin down are degenerate. Standard calculation\cite{Coleman2015} gives
\be
\Pi_0(i\nu,q) = -\lp \nu_0 (1-l_0^2q^2) - \frac{|\nu|}{q}\frac{p_1+p_2}{4\pi v_F^2}\rp
\ee
where $\nu_0$ is the density of state at the Fermi level,  $p_1$ and $p_2$ are the radii of curvature on the Fermi surface where $q$ is perpendicular to the fermi velocity. The quantity $l_0$ is a length scale depending on band dispersion. 

Importantly, the momentum $q$ relevant for analysis is restricted to a certain direction. Namely, the momentum $q$, which is the momentum transfer of the interaction used to generate CDW, is parallel to the tangential of the Fermi surface at $Q$(see main text), where the curvature of radius is maximized. Therefore, for such a $q$, one of $p_1$ and $p_2$ is $p_*$, which is the radius of curvature at $Q$ point, whereas the other is $p_*'$ which is the radius of curvature at the point opposite to $Q$ point on the Fermi surface. This definition of $p_*$ and $p_*'$ is the same as the definition in the main text. Therefore, the polarization function at the relevant $q$ can be written as
\be
\Pi_0(i\nu,q) = -\lp \nu_0 (1-q^2l_0^2) - \frac{|\nu|}{q}\frac{p_*+p_*'}{4\pi v_F^2}\rp
\ee
Using the Stoner condition $\nu_0 U=1$, and plugging in the density of state in our model $\nu_0=m/2\pi$, we find
\be
\gamma(i\nu,q) = \frac{1}{ \frac{\kappa_*+\kappa'}{2}\frac{|\nu|}{v_Fq} + q^2l_0^2 }
\ee
Further accounting for the screening effect, we arrive at the following form of effective coupling
\be\label{eq:sup_V}
V(i\nu,q) = \frac{U\gamma(i\nu, q)^2}{1-NU\gamma(i\nu, q)\Pi_0(i\nu, q)} \sim U\gamma(i\nu, q)/N = \frac{U/N}{ \frac{\kappa_*+\kappa'}{2}\frac{|\nu|}{v_Fq} + q^2l_0^2 }
\ee
where $N=4$ represents the number of isospin species in the unpolarized phase.
Eq.\eqref{eq:sup_V} is the form we used in the main text Eq.\eqref{eq:H_eff1}. Here we have used the stoner condition $-U\Pi_0(0,0)=1$ and that each polarization bubble is renormalized by one vertex correction $\gamma(i\nu,q)$, which is divergent at small momentum and low frequency. 

\section{ Renormalization of the CDW/SDW vertex correction}
	In this section, we derive the first-order correction for CDW/SDW vertex function, which is a result used in main text Eq.\eqref{eq:1st-order_vertex_correction}. This derivation is similar to the analysis of the gauge-fluctuation problem carried out in Ref.\cite{Altshuler1994}. We start from the following vertex correction which is read off from the diagram in main text Fig.\ref{fig:vertex_correction} a):
	\be\label{eq:self-consistency_2}
	\delta \Gamma(\epsilon) = - \sum_{\nu, \vec k}\frac{V(i\nu-i\epsilon,k)\Gamma_0(\nu)}{\lp i\nu-\frac{k_\perp^2}{2m_*}\rp^2-v_F^2k_\parallel^2}
	\ee
	To proceed analytically, the following approximation can be employed. First, 
	we focus on the frequency regime $\nu\gg \epsilon$ which , as we will see, is the origin of log-divergence. In this regime, $V(i\nu-i\epsilon)$ can be substituted with $V(i\nu)$. Second, 
	the contribution to the right-hand side decays much faster along $k_\parallel$ direction than along $k_\perp$ direction. As a result, processes with $k_\perp\gg k_\parallel$ dominates the right-hand side. 
	Therefore, we can replace $|k|$ in the expression of $V(i\nu,k)$ with $k_\perp$. After that, Eq.\eqref{eq:self-consistency_2} becomes
	\begin{align}
		& \delta\Gamma (\epsilon) = - \frac{4Uv_F}{N\kappa } \int_\epsilon \frac{d\nu}{2\pi} \sum_{k_\perp} \frac{|k_\perp|}{\lp i\nu-\frac{k_\perp^2}{2m_*}\rp^2-v_F^2k_\parallel^2} \frac{1}{|\nu|+\alpha |k_\perp|^3} \Gamma_0,\label{eq:self-consistency_2_approx}
	\end{align}
	The competition of two denominators in the integral define a new energy scale.
	Namely, the integrand is suppressed by the first factor at a momentum scale of $k_\perp \gtrsim k_1 = \epsilon^{1/3} \alpha^{-1/3}$ whereas the second factor starts to suppress the integrand at $k_\perp \gtrsim k_2= (2m\epsilon)^{1/2}$. The comparison of these two momentum scales $k_1$ and $k_2$ sets a characteristic frequency 
	\be 
	\epsilon_* = \alpha^{-2}(2m)^{-3}.
	\ee 
	which is set by the details in band dispersion. From now on we focus on the case where $\epsilon_*>\epsilon_F$ because it simplifies the discussion as we always have $\epsilon<\epsilon_*$ in this case. In this regime, $k_1>k_2$, so the $k_\perp^3$ term in the denominator can be safely neglected, yielding
	\be\label{eq:deltaGamma_2}
		\delta\Gamma (\epsilon) = - \frac{4Uv_F}{ N\kappa }\int^{\epsilon_F}_\epsilon \frac{d\nu}{2\pi} \frac{S(\nu)}{|\nu|}\Gamma_0,\quad
		S(\nu)=\sum_{k_\perp} \frac{|k_\perp|}{\lp i\nu-\frac{k_\perp^2}{2m_*}\rp^2-v_F^2k_\parallel^2}
	\ee

	We 
	consider exclusively the case where $\Gamma(\epsilon)$ is even in $\epsilon$ since the odd-frequency channel is weak due to an extra zero at $\epsilon=0$. 
	Therefore, we define a quantity $S'$ to describe the symmetrized part of $S$:
	\be
	S'(\nu)=\frac{S(\nu) + S(-\nu)}{2} 
	\ee
	Below, we evaluate $S'$ by working out the summation over $k$ step by step. First, we integrate over $k_\parallel$ to obtain 
	\be
	S'(\nu) =  {\rm Re} \lb -\frac{i \sgn(\nu)}{2v_F}\sum_{k_\perp}\frac{|k_\perp|}{ i\nu-\frac{k_\perp^2}{2m_*} }\rb
	\ee
	Next, we integrate over $k_\perp$, finding
	\begin{align}
		S'(\nu)  &= {\rm Re}\lb -\frac{i\sgn(\nu)}{v_F} 2\int _0^{\infty} \frac{k_\perp dk_\perp}{ 2\pi \lp i\nu-\frac{k_\perp^2}{2m_*} \rp}\rb\nonumber\\
		&=-\frac{i\sgn(\nu) 2m_*}{2\pi v_F} \int _0^{\infty} \frac{-4im_*\nu   k_\perp dk_\perp}{  \lp 2m_*\nu\rp ^2 + k_\perp^4 }\nonumber\\
		&= -\frac{2m_*}{2\pi v_F} \frac{\pi}{2} = -\frac{m_*}{2 v_F}\label{eq:S'(nu)} 
	\end{align}
	Plugging back into Eq.\eqref{eq:deltaGamma_2} we find
	\be \label{eq:self-consistency_log}
	\delta\Gamma(\epsilon) = \frac{2Um_*}{ N\kappa} \int^{\epsilon_F} _\epsilon \frac{d\nu}{2\pi\nu} \Gamma_0 =  \frac{Um_*}{\pi N\kappa}   \ln\lp \frac{\epsilon_F}{\epsilon} \rp \Gamma_0.
	\ee
	This is the result used in main text Eq.\eqref{eq:1st-order_vertex_correction}. Interestingly, we arrive at the same logarithmically divergent vertex correction as in the problem of Fermi sea coupled by gauge-field fluctuation studied in Ref.\cite{Altshuler1994}, despite the absence of singular self-energy in our analysis. The presence or absence of self-energy does not affect the logarithmic divergence because the self-energy ultimate goes into $S(\nu)$, in which the integral over $k_\perp$ always yields a constant as shown in Eq.\eqref{eq:S'(nu)}, so that the self-energy drops out.

\end{document}